\documentclass[lineno]{jfm}
\usepackage{graphicx}
\usepackage{bm}
\usepackage{graphicx}
\usepackage{amsmath}
\usepackage{amssymb}
\usepackage{pifont}
\usepackage{epstopdf}

\title{Reciprocal theorem for ion-releasing colloidal particles. }

\author{Evgeny S. Asmolov\aff{1} \and Olga I. Vinogradova\aff{1}}

\affiliation{\aff{1}Frumkin Institute of Physical Chemistry and Electrochemistry,
Russian Academy of Sciences, 31-4 Leninsky Prospect, 119071 Moscow, Russia}

\begin{document}
\maketitle

\begin{abstract}
We present a generalization of the Lorentz reciprocal theorem for particles suspended in electrolyte solutions and subjected to an electric field, which can be either externally applied or spontaneously generated. Attention is focused on catalytic colloids that actively release ions into the solution. The primary advantage of this generalization is its ability to capture the effect of a secondary diffuse cloud forming around the catalytic particle, which is equivalent to accounting for the net excess charge $Q$ of the system. Our results show that the phoretic propulsion speed of such particles features an extra contribution proportional to $Q$ and the far-field electric field $E_{\infty}$. The derived analytical expression for $Q$ reveals that its sign is governed by the difference in ionic diffusivities, while its magnitude is controlled solely by the average surface flux of ions. We demonstrate the application of the generalized theorem to the electrophoresis and diffusiophoresis of passive particles characterized by a homogeneous ionic emission, as well as to the self-propulsion of active anisotropic microswimmers. It is shown that while in certain configurations the extra term vanishes or has a negligible effect, in many other regimes it can dramatically alter both the magnitude and the sign of the phoretic mobility. Finally, the critical relevance of our findings for long-range microswimmer interactions and emergent collective behavior is highlighted.
\end{abstract}

%\begin{keywords}
%Authors should not enter keywords on the manuscript, as these must be chosen by the author during the online submission process and will then be added during the typesetting process (see \href{https://www.cambridge.org/core/journals/journal-of-fluid-mechanics/information/list-of-keywords}{Keyword PDF} for the full list).  Other classifications will be added at the same time.
%\end{keywords}

\section{Introduction}\label{sec:intro}

The Lorentz reciprocal theorem represents a powerful analytical tool in low-Reynolds-number hydrodynamics \citep{lorentz1897,happel2012low}, which is successfully used to solve a variety of complex physical problems. This theorem provides a direct integral relation between the main problem to be solved and a much simpler auxiliary problem with a known solution. In other words, it allows one to bypass solving the explicit, full boundary-value problems for velocity and pressure fields, enabling the direct calculation of macroscopically observable quantities such as forces, torques, or particle propulsion speeds \citep{masoud2019}. Over the decades, the reciprocal theorem has been successfully employed to investigate many different systems, ranging from the classical sedimentation of particles to their autophoretic migration relative to a solvent \citep{teubner1982,brady2011,asmolov2022COCIS,hosaka.y:2023,ganguly2024}.

The migration of particles driven by a salinity gradient or an external electric field, termed diffusiophoresis and electrophoresis, respectively, has played a very great part in the development of colloid science and remains the subject of active research \citep{shim2022,ault2025}. This enduring interest is largely due to the fact that the controlled manipulation of microparticles is a key issue for a wide variety of modern technologies, including lab-on-a-chip systems and targeted drug delivery \citep{moran2017,hu2020micro,ju2025technology}. While extensive theoretical efforts have historically focused on the phoretic propulsion of inert passive particles \citep{deryagin1961,overbeek}, the focus of contemporary soft matter physics has significantly shifted toward active catalytic particles that generate ionic fluxes through surface chemical reactions. A quantitative understanding of their migration, however, remains challenging because it requires a simultaneous description of the fluid flow, electrostatics, and ionic transport, which are all strongly entangled near the charged surface \citep{moran2017}.

In this context, utilizing the reciprocal theorem would be highly preferable to directly solving the full system of coupled partial differential equations, especially for heterogeneous particles or those of arbitrary shapes. Consequently, there is a growing literature describing attempts to apply this theorem to evaluate the propulsion speed of both passive catalytic colloids migrating only in external fields \citep{asmolov.es:2024,asmolov2024electro} and active self-propelling microswimmers \citep{ibrahim2017,de2020self,asmolov2022MDPI,asmolov2022self,xiao2025ionic}. However, the assumption that the classical formulation of the reciprocal theorem holds unconditionally for catalytic particles is by no means obvious. The point is that near such particles, in addition to the standard compensating electric double layer, a secondary, macroscopic and weakly charged ionic cloud typically emerges. It is therefore natural to enquire whether this extended cloud affects the familiar formulation of the theorem — a fundamental issue that has been completely ignored so far.

In this paper, we propose a generalization of the reciprocal theorem to the case of catalytic particles. In \S\ref{sec:general}, we describe the underlying model and formulate the governing electro-hydrodynamic relationships. An equation relating the particle velocity to the macroscopic excess charge of the system is then derived in \S\ref{sec:velocity}, followed by the explicit calculation of this charge in \S\ref{sec:charge}. We close in \S\ref{sec:examples} by demonstrating the applicability of our framework to both passive and active particles, and offering further outlook. Finally, concluding remarks and potential extensions are summarized in \S\ref{sec:conclusion}.

\section{General consideration}\label{sec:general}
\subsection{Model}

We consider a colloid particle immersed in an aqueous salt solution of concentration $C_{\infty}$ subject to an electric field  in the $x-$direction. In many cases, it is observed to migrate relative to the solvent with a velocity $\mathbf{V}_{p}$ that has to be found. \\

The particle can be (chemically) \emph{inert} as in most classical studies~\citep{smoluchowski.m:1921,overbeek,prieve.dc:1984}. In a solution of electrolyte, it assumes a certain charge. A compensating charge of the opposite sign and equal magnitude is staying in the electrostatic diffuse layer (EDL) in the neighborhood of the  inert surface, so the total charge $Q$ of the region outside the EDL is zero. The extension of this EDL is of the same
order as the Debye length, $\lambda_D = \left( 8\pi e^{2}C_{\infty }/\left(
\epsilon k_{B}T\right) \right) ^{-1/2}$ of the solution, where $e$ is the elementary positive charge, $\epsilon$ is permittivity, $k_{B}$ is the
Boltzmann constant, and $T$ is the temperature.  When the field is applied such a particle moves with a constant speed $\mathbf{V}_{p}$ since volume elements of the liquid around  are subjected to two equal forces of opposite directions, one due to the electric field working on the charge in the EDL, the other due to viscous friction. \\

\begin{figure}[ht]
\centering
\includegraphics[width=0.8\linewidth]{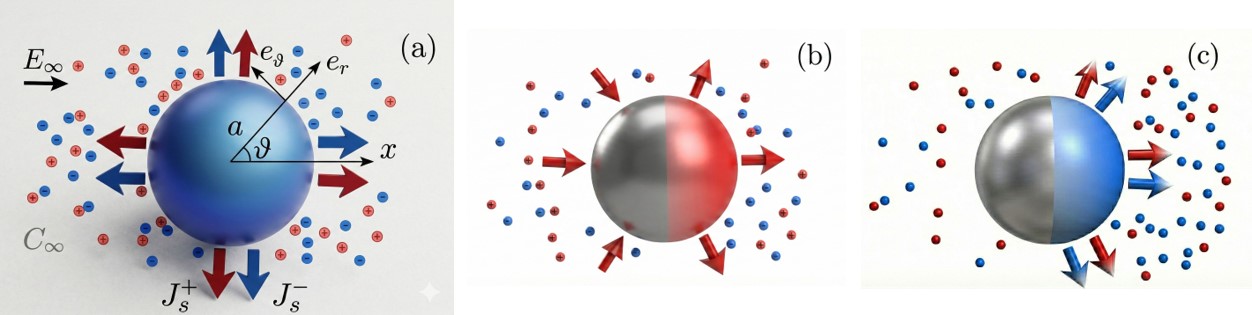}
\caption{Sketch of the passive catalytic particle (a) and of the active particles or swimmers of Types I (b) and II (c) releasing inhomogeneous fluxes of ions.}
\label{fig1}
\end{figure}

Here we focus on the ion-releasing particles termed \emph{catalytic}. If such a release is uniform as depicted in Fig.~\ref{fig1}(a), the particle remains \emph{passive} at small P\'{e}clet numbers ($Pe \ll 1$), well below the threshold for spontaneous symmetry breaking \citep{michelin2013}, and can only migrate under an applied external electric field $E_{\infty }$. However, when ions are released non-uniformly, a different story obtains. The particle becomes \emph{active}, that is, capable of self-propelling. Such particles termed swimmers can be broadly classified into two different categories depending on the nature of the ionic flux from their surface. The ones that release solely one type of ion from the active part with their simultaneous absorption by another side, as illustrated in Fig.~\ref{fig1}(b), are of Type I. For swimmers of Type II the active sectors release both cations and anions, whilst another part of their surface is inert as shown in Fig.~\ref{fig1}(c). In the case of active particles an electric field emerges spontaneously (but can, of course, also be applied externally). \\

\begin{figure}[ht]
\centering
\includegraphics[width=0.8\linewidth]{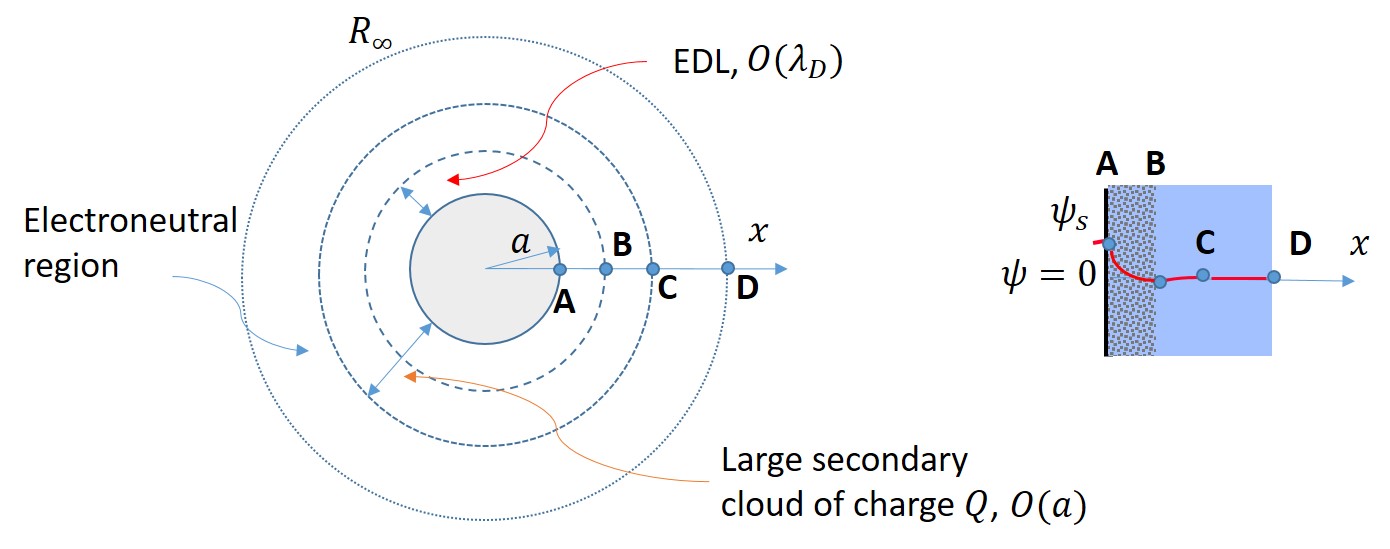}
\caption{Schematic plot of the ionic clouds around the passive catalytic particle and corresponding potential distribution. The circles mark the boundaries of regions.}
\label{fig1b}
\end{figure}

Before turning to the equations, it is instructive to ask what is unique about catalytic particles dispersed in electrolyte solutions.
The conventional theories of phoretic propulsion typically treat the fluid outside the thin electric double layer (EDL) as locally electroneutral. However, this is not, in general, the case when ions are actively released.
When a catalytic particle, releasing ions with distinct diffusivities ($D^+ \neq D^-$), is placed in an electrolyte, the outward diffusion fluxes across an imaginary sphere of a large but finite radius $R_{\infty}$ differ initially. During the transient period, a fraction of the faster ion species escapes beyond that sphere and disperses into the far-field bulk. Due to this escape of charge by fast ions, an excess net charge of the opposite sign accumulates within the sphere. This accumulation grows until the emerging macroscopic electric field equilibrates the ion fluxes, slowing the faster ions and accelerating the slower ones to establish a steady state.

This mechanism implies that in the steady state, besides the familiar thin EDL, a macroscopic, weakly charged cloud is spontaneously formed~\citep{asmolov2022self,asmolov.es:2024}. This secondary cloud, characterized by a self-consistently established electrostatic potential $\Psi$, extends to distances of $O(a)$, as schematically illustrated in Fig.~\ref{fig1b}. If such a cloud exists—and mounting evidence suggests it must—it clearly has profound implications for understanding the particle propulsion speed. The key physical point is that while the EDL (the region between $\mathbf{A}$ and $\mathbf{B}$), though distorted by the released ions, balances the surface charge of the particle, the large bulk cloud between $\mathbf{B}$ and $\mathbf{C}$ carries a net volume charge. Because the compensating fast ions have escaped beyond the boundary $\mathbf{D}$ (of radius $R_{\infty}$), the total integrated charge inside it takes on a non-zero value. This net charge $Q$ becomes a crucial consideration in determining the propulsion speed; however, existing theories fail to accommodate it.

Our purpose in this paper is to generalize the reciprocal theorem for catalytic particles by explicitly incorporating this missing charge contribution and to illustrate its relevance for the situations shown in Fig.~\ref{fig1}.

\subsection{Summary of general electro-hydrodynamic relationships}

We describe the system using a resting spherical coordinate system $\left( R,\theta ,\varphi \right)$ centered on the particle of radius $a$. The fluid flow is governed by the Stokes equations incorporating the electrical body force:
\begin{eqnarray}
\mathbf{\nabla \cdot U} &=& 0, \label{cont} \\
\mathbf{\nabla \cdot \boldsymbol{\sigma} }_{h}+\mathbf{\nabla \cdot \boldsymbol{\sigma} }_{e} &=& \eta \nabla^2 \mathbf{U}-\mathbf{\nabla }P+\rho \mathbf{E} = \mathbf{0},  \label{mom}
\end{eqnarray}
where $\mathbf{U}$ and $P$ denote the fluid velocity and pressure, $\eta$ is the dynamic viscosity, $\rho$ represents the volume electric charge density, and $\mathbf{E}$ is the electric field. The hydrodynamic and electrostatic stress tensors, $\boldsymbol{\sigma}_{h}$ and $\boldsymbol{\sigma}_{e}$, are defined conventionally as:
\begin{equation*}
\boldsymbol{\sigma }_{h}=-P\mathbf{I}+\frac{1}{2}\left[ \mathbf{\nabla U}+\left(\mathbf{\nabla U}\right)^{T}\right], \quad
\boldsymbol{\sigma }_{e}=\frac{\epsilon }{4\pi }\left( \mathbf{EE}-\mathbf{I}\frac{E^{2}}{2}\right),
\end{equation*}
with $\mathbf{I}$ being the unit tensor. At the particle surface ($R=a$), the fluid satisfies the standard no-slip boundary condition: $\left. \mathbf{U}\right\vert _{R=a}=\mathbf{V}_{p}$.

The total force acting on the particle is the sum of the electrostatic and hydrodynamic contributions, which represent the integrals of the corresponding stresses over the particle surface $S_{p}$:
\begin{equation}
\mathbf{F}_{e}=\int_{S_{p}}\mathbf{\sigma }_{e}\cdot \mathbf{n}dS,\quad
\mathbf{F}_{h}=\int_{S_{p}}\mathbf{\sigma }_{h}\cdot \mathbf{n}dS,
\label{force}
\end{equation}
where $\mathbf{n}$ is the outward unit normal to the surface. In the steady state, the total force on the particle is zero:
\begin{equation}\label{eq:zeroF}
 \mathbf{F}_{e} + \mathbf{F}_{h} =\mathbf{0}.
\end{equation}

To apply the reciprocal theorem to our system, we introduce an auxiliary purely hydrodynamic flow past an inert, uncharged particle of identical radius translating at a unit velocity $\mathbf{e}_{p}$ in an arbitrary direction. The corresponding auxiliary Stokes flow satisfies:
\begin{equation}
\mathbf{\nabla \cdot \widehat{U}}=0,\quad \mathbf{\nabla }\cdot \boldsymbol{\widehat{\sigma }}_{h}=\mathbf{0},  \label{St}
\end{equation}
subject to the no-slip condition $\left. \mathbf{\widehat{U}}\right\vert _{R=a}=\mathbf{e}_{p}.$ The velocity field for this reference problem can be expressed via the standard resistance tensor \citep{happel2012low,stone2016}:
\begin{equation*}
\widehat{\mathbf{U}}=\mathbf{e}_{p}\cdot \mathbf{D},
\end{equation*}
where the tensor $\mathbf{D}$ takes the explicit form:
\begin{equation*}
\mathbf{D}=\frac{3a}{4R}\left( \mathbf{I}+\frac{\mathbf{RR}}{R^{2}}\right) +
\frac{3a^{3}}{4R^{3}}\left( \mathbf{I}-3\frac{\mathbf{RR}}{R^{2}}\right).
\end{equation*}
For this auxiliary configuration, the total hydrodynamic force acting on the particle simplifies to the classical Stokes drag, $\mathbf{\widehat{F}}_{h}=-6\pi \eta a\mathbf{e}_{p}.$

For systems carrying a space charge, the reciprocal theorem formulated by \citet{teubner1982} takes the form:
\begin{equation}
\int_{S_{p}}\mathbf{n}\cdot \boldsymbol{\sigma}_{h}\cdot \mathbf{\widehat{U}}
dS-\int_{V}\rho \mathbf{\widehat{U}}\cdot \mathbf{E}dV=\int_{S_{p}}\mathbf{n}\cdot \boldsymbol{\widehat{\sigma }}_{h}\cdot \mathbf{U}dS,  \label{rec1}
\end{equation}
where $V$ denotes the fluid volume. Because the fluid velocity at the particle surface matches the rigid-body translation, these surface integrals can be directly re-expressed in terms of the total electro-hydrodynamic forces:
\begin{equation}
\int_{S_{p}}\mathbf{n}\cdot \boldsymbol{\sigma}_{h}\cdot \mathbf{\widehat{U}}dS=\mathbf{F}_{h}\cdot \mathbf{e}_{p},\quad \int_{S_{p}}\mathbf{n}\cdot \boldsymbol{\widehat{\sigma }}_{h}\cdot \mathbf{U}dS=-6\pi \eta a\mathbf{e}_{p}\cdot\mathbf{V}_{p}.
\label{maxw}
\end{equation}
Since every term in \eqref{rec1} is linearly proportional to the arbitrary unit vector $\mathbf{e}_{p}$, this projection can be factored out, yielding the standard relation for the electro-hydrodynamic force \citep{teubner1982,ganguly2024}:
\begin{equation}
\mathbf{F}_{h}=-6\pi \eta a\mathbf{V}_{p}+\int_{V}\rho \mathbf{D\cdot E}dV.  \label{fh}
\end{equation}

Once the electrostatic force is determined, Eqs.~\eqref{eq:zeroF} and \eqref{fh} provide a direct route to the particle velocity. We turn now to its calculation.

\section{Evaluation of the phoretic velocity}\label{sec:velocity}

For an inert particle, the derivation of the governing velocity relation is straightforward. The net electrostatic force acting on the colloid is given by $\mathbf{F}_{e}=-\int_{V}\rho \mathbf{E}dV$, where the negative sign arises because the outward unit normal $\mathbf{n}$ points toward the interior of the integrated fluid volume $V$. By substituting this force into the global force balance \eqref{eq:zeroF} and utilizing relation \eqref{fh}, we find that the particle velocity takes the explicit form:
\begin{equation}
\mathbf{V}_{p}=-\frac{1}{6\pi \eta a} \int_{V}\rho \left( \mathbf{D}-\mathbf{I}\right) \mathbf{\cdot E}dV.
\label{rec0}
\end{equation}

The extension of this approach to a catalytic particle, however, requires careful handling due to the net volume charge accumulated within the secondary cloud. To determine the electrostatic force $\mathbf{F}_{e}$ in this case, we apply Gauss's divergence theorem to the Maxwell stress tensor $\boldsymbol{\sigma}_{e}$:
\begin{equation}
\int_{S}\boldsymbol{\sigma}_{e}\cdot \mathbf{n}dS=-\int_{V}\rho \mathbf{E}dV,
\label{gauss_e}
\end{equation}
where $S=S_{p}+S_{\infty}$ is the total boundary enclosing the fluid volume $V$. Then, by using \eqref{force}, we explicitly determine the electrostatic force acting on the particle:
\begin{equation}
\mathbf{F}_{e}=-\int_{V}\rho \mathbf{E}dV-\int_{S_{\infty }}\boldsymbol{\sigma}_{e}\cdot \mathbf{n}dS.  \label{gauss}
\end{equation}

The standard (inert) term is thus supplemented by an electrostatic boundary contribution evaluated over the control surface $S_{\infty}$ (a sphere of radius $R \gg a$ enclosing the system). We emphasize that no such extra terms appear in the hydrodynamic contributions to the reciprocal theorem \eqref{rec1}. Because the hydrodynamic fields decay rapidly in the far field—with the velocity scaling as $R^{-1}$ and the viscous stress as $R^{-2}$—the corresponding surface integrals over $S_{\infty} = 4\pi R^{2}$ vanish asymptotically as $O(a R^{-1}) \ll 1$. For the electrostatic integral, however, this scaling does not hold. Indeed, the electric field far from the particle is given by:
\begin{equation}
\mathbf{E}=E_{\infty }\mathbf{e}_{x}+\frac{Q}{\epsilon R^{2}}\mathbf{e}_{r}+O\left( R^{-3}\right).  \label{efar}
\end{equation}
The integration over $S_{\infty }$ then reduces exactly to:
\begin{equation}
\int_{S_{\infty }}\boldsymbol{\sigma}_{e}\cdot \mathbf{n}dS=-E_{\infty }Q\mathbf{e}_{x},  \label{co3}
\end{equation}
which, when combined with \eqref{gauss}, yields the total electrostatic force acting on the particle:
\begin{equation}
\mathbf{F}_{e}= - \int_{V}\rho \mathbf{E}dV+E_{\infty }Q\mathbf{e}_{x}.
\label{fe}
\end{equation}

While the first term in \eqref{fe} can point in an arbitrary direction depending on the cloud distortion, the second contribution remains strictly aligned with the external electric field. Physically, this extra term represents the net Coulomb force exerted by the external field on the charged secondary cloud, disappearing naturally in the absence of an applied field or when the net system charge vanishes ($Q = 0$).

Using Eqs.~\eqref{eq:zeroF}, \eqref{fh}, and \eqref{fe}, we then easily obtain the generalized equation for the particle velocity:
\begin{equation}
\mathbf{V}_{p}=\frac{1}{6\pi \eta a}\left[ \int_{V}\rho \left( \mathbf{D} - \mathbf{I}\right) \mathbf{\cdot E}dV+E_{\infty }Q\mathbf{e}_{x}\right].  \label{rec}
\end{equation}
Below we calculate the total charge $Q$, which remains to be determined.

\section{Calculation of the total system charge}\label{sec:charge}

In this section, we calculate the total excess charge $Q$ by invoking a standard asymptotic assumption in classical linear electrokinetics. Specifically, we assume that the external electric field $E_{\infty }$ is weak compared to the intrinsic radial field induced spontaneously by the catalytic particle itself. This scale separation ensures that the external field does not alter the net charge distribution to the leading order, meaning that $Q$ is induced solely by the particle's own ionic release.

Under these steady-state conditions, the conservation of each ionic species at every point in the fluid volume is governed by the Nernst–Planck transport equation:
\begin{equation}
\mathbf{\nabla }\cdot \mathbf{J}^{\pm }=0,  \label{NP}
\end{equation}
where the corresponding ionic fluxes $\mathbf{J}^{\pm }$ are defined as:
\begin{equation}
\mathbf{J}^{\pm }=D^{\pm }\left( -\mathbf{\nabla }C^{\pm }\mp \dfrac{e}{k_{B}T}C^{\pm }\mathbf{\nabla }\Psi \right).  \label{jd}
\end{equation}
Here, the upper and lower signs denote the positive and negative ions, respectively, $C^{\pm }$ represent the ion number densities, $D^{\pm }$ are the respective ion diffusivities, and $\Psi $ is the self-consistent electrostatic potential. The first term in \eqref{jd} accounts for the classic diffusive drift relative to the solvent, whereas the second term describes the migration of ions driven by the emerging macroscopic electric field. We note that the convective contributions to the total fluxes are neglected here, which remains fully justified in the limit of small P\'eclet numbers ($Pe \ll 1$).

The self-consistent electrostatic potential $\Psi$ and the local charge density $\rho$ are coupled via the Poisson equation:
\begin{equation}
\nabla^2 \Psi =-\frac{4\pi \rho }{\epsilon }=-\frac{4\pi e\left(C^{+}-C^{-}\right)}{\epsilon}.  \label{PEq}
\end{equation}

At the particle surface, the boundary conditions prescribe the specific emission fluxes of the released ions:
\begin{equation}
\mathbf{R}\in S_{p}:\ \mathbf{J}^{\pm }\cdot \mathbf{n}=J_{s}^{\pm }\left(\mathbf{R}\right).  \label{sp2}
\end{equation}
Strictly speaking, the long-term stability of a steady state dictates that the net electrical charge of the colloid must remain constant. This conservation requirement demands a vanishing total current, meaning that the integrated surface fluxes of cations and anions must balance each other perfectly:
\begin{equation}
\int_{S_{p}}J_{s}^{+}dS=\int_{S_{p}}J_{s}^{-}dS=4\pi a^{2}\left\langle J\right\rangle,
\label{norm0}
\end{equation}
where $\left\langle J\right\rangle$ represents the average ionic flux at the surface. In the far field, the system relaxes back to its uniform equilibrium state:
\begin{equation}
R\rightarrow \infty :\ C^{\pm }=C_{\infty }, \quad \Psi =0. \label{bcc}
\end{equation}

Crucially, the ionic transport problem \eqref{NP}–\eqref{bcc} decouples entirely from the fluid mechanics. Consequently, the spatial distribution of the ionic fields remains independent of the phoretic velocity $\mathbf{V}_{p}$ and can be determined prior to solving the hydrodynamic problem.

Applying the divergence theorem to Eq. (\ref{NP}) yields a global balance of ionic species:
\begin{equation}
\int_{V}\mathbf{\nabla }\cdot \mathbf{J}^{\pm }dV=-\int_{S_{p}}\mathbf{J}^{\pm }\cdot \mathbf{n}dS-\int_{S_{\infty }}\mathbf{J}^{\pm }\cdot \mathbf{n}dS=0,  \label{gauss2}
\end{equation}
where the first integral on the right-hand side evaluates to $4\pi a^{2}\left\langle J\right\rangle$. The second boundary integral, reflecting the fluxes at infinity, can be calculated by integrating the Poisson equation (\ref{PEq}) over the macroscopic volume to yield:
\begin{equation}
\int_{S_{\infty }}\mathbf{\nabla }\Psi \cdot \mathbf{n}dS=-\frac{4\pi Q}{\epsilon }.  \label{gauss3}
\end{equation}

To determine this term explicitly, we invoke the far-field asymptotic behavior of the ionic fluxes defined in Eq.~\eqref{jd}. Assuming that the concentrations and potential decay as $R^{-1}$ far from the particle, we can express them via the boundary conditions \eqref{bcc} as:
\begin{equation}
C^{\pm }=C_{\infty }+\frac{aC_{-1}^{\pm }\left( \theta ,\varphi \right) }{R}+O\left( R^{-2}\right), \quad \Psi =\frac{a\Psi _{-1}\left( \theta ,\varphi \right) }{R}+O\left( R^{-2}\right).
 \label{cex}
\end{equation}

Crucially, Eq.~\eqref{PEq} dictates that the volume charge density must scale as $\rho \propto \nabla ^2 \Psi \propto \Psi _{-1}/R^{3}$. This faster decay of the net charge density relative to the individual ion concentrations ($C^{\pm }$) requires that the leading-order angular perturbations in the expansion \eqref{cex} must balance each other, giving:
\begin{equation}
C_{-1}^{+}=C_{-1}^{-}=C_{-1}\left( \theta ,\varphi \right).  \label{cex1}
\end{equation}

Consequently, the far-field concentrations of anions and cations remain equal up to $O\left(R^{-2}\right)$. Substituting these expansions into the Nernst–Planck equations (\ref{NP}) and collecting the leading-order terms, we find that the transport simplifies to:
\begin{equation}
- \nabla ^2 \left( \frac{aC_{-1}\left( \theta ,\varphi \right) }{R}\right) \mp \dfrac{e}{k_{B}T}C_{\infty }\nabla ^2 \left( \frac{a\Psi _{-1}\left( \theta ,\varphi \right) }{R}\right) =0.
\end{equation}

By sequentially summing and subtracting these expressions, we conclude that both the concentration and the potential perturbations satisfy the Laplace equation:
\begin{equation}
\nabla ^2 \left( \frac{C_{-1}\left( \theta ,\varphi \right) }{R}\right) =0,\quad \nabla ^2 \left( \frac{\Psi _{-1}\left( \theta ,\varphi \right) }{R}\right) =0.  \label{lap2}
\end{equation}

Since any solution to the Laplace equation decaying as $R^{-1}$ must be spherically symmetric, both $C_{-1}$ and $\Psi _{-1}$ reduce to constants independent of $\theta$ and $\varphi $. This allows us to explicitly evaluate the far-field flux integral in Eq.~\eqref{gauss2}:
\begin{equation}
\int_{S_{\infty }}\mathbf{J}^{\pm }\cdot \mathbf{n}dS=4\pi D^{\pm }a\left(C_{-1}\pm \dfrac{eC_{\infty }}{k_{B}T}\Psi _{-1}\right) =-4\pi a^{2}\left\langle J\right\rangle.
\end{equation}

Finally, enforcing the condition that the total fluxes of anions and cations are equal yields the relationship:
\begin{equation}
\Psi _{-1}=-\beta \dfrac{k_{B}T}{eC_{\infty }}C_{-1}=\beta \dfrac{ak_{B}T}{eC_{\infty }D}\left\langle J\right\rangle,
\label{psi1}
\end{equation}
where the diffusivity contrast $\beta$ and the effective diffusion coefficient $D$ are defined as:
\begin{equation}
\beta =\frac{D^{+}-D^{-}}{D^{+}+D^{-}},\quad D=\frac{2D^{+}D^{-}}{D^{+}+D^{-}}.  \label{bet}
\end{equation}

Equation (\ref{psi1}) is equivalent to the well-known condition of zero current in the electroneutral region for a planar geometry \citep{deryagin1961,prieve.dc:1984},
\[
\mathbf{\nabla }\Psi =-\beta \dfrac{k_{B}T}{e}\mathbf{\nabla }\left( \ln C\right),
\]
where constant gradients do not require the emergence of a volume charge. However, for a spherical geometry, the gradients are inhomogeneous in the far region ($\propto 1/R^2$), which fundamentally implies a finite total charge of the system within the finite boundary $R_{\infty}$.

By recognizing that the integral evaluation reduces to the relation $a\Psi_{-1} = -Q/\epsilon$, we can directly substitute this result into \eqref{rec} to obtain the final explicit expression for the total excess charge of the system:
\begin{equation}
Q=-\epsilon \beta \dfrac{a^{2}k_{B}T}{eC_{\infty }D}\left\langle J\right\rangle.
\label{qdim}
\end{equation}

This result clearly demonstrates that the net charge $Q$ is entirely independent of the intrinsic surface charge or electrostatic potential of the colloid; instead, it is driven solely by the average ionic surface flux $\left\langle J\right\rangle$, while its sign is governed strictly by the diffusivity contrast $\beta$. In practice, since the parameter $\beta$ is negative for common inorganic sodium and lithium salts, their presence induces a positive net system charge $Q$. Conversely, potassium salts exhibit a virtually vanishing diffusivity contrast ($\beta \simeq 0$), which fundamentally suppresses this effect and causes the total excess charge to approach zero ($Q \rightarrow 0$).

\section{Examples and outlook}\label{sec:examples}

\subsection{Electrophoresis and diffusiophoresis of passive particles}

To demonstrate the physical implications of the developed framework, we first apply our generalized formulation to the classic phenomena of electrophoresis and diffusiophoresis of a passive catalytic particle. This system is characterized by a stoichiometrically balanced ionic emission flux, $J_{s} = J_{s}^{+} = J_{s}^{-}$, from its surface, as schematically illustrated in Fig.~\ref{fig1}(a). The existing theories for these phenomena, including our recent calculations \citep{asmolov.es:2024,asmolov2024electro}, relied on the conventional application of the reciprocal theorem. Within that standard paradigm, the emergence of a macroscopic excess charge was overlooked, and consequently, the additional phoretic term associated with $Q \neq 0$ could not be accommodated in the velocity relation \eqref{rec}. The generalized approach developed here explicitly resolves this fundamental limitation.

Electrophoresis refers  to the migration of a particle under a constant applied electric field, which is conventionally assumed to be weak, $\mathcal{E} = E_{\infty }ea/\left( k_{B}T\right) \ll 1$. Conversely, diffusiophoresis represents the migration driven by an imposed concentration gradient, which is likewise treated as small, such that $C=C_{\infty }\left( 1+\mathcal{E}X/a\right)$ at $R_{\infty}$. In the latter case, the macroscopic electric field $\mathcal{E}\beta k_{B}T/\left( ea\right)$ is not applied externally; rather, it emerges spontaneously to ensure a vanishing total current density far from the sphere \citep{prieve.dc:1984}.

Let us introduce the dimensionless variables:
\begin{equation}
\mathbf{r} = \frac{\mathbf{R}}{a}, \quad c=\frac{C}{C_{\infty}}, \quad \psi =\dfrac{e\Psi }{k_{B}T},\quad
\mathbf{v}_p=\mathbf{V}_p\dfrac{4\pi \eta e^{2}a}{\epsilon k_{B}^{2}T^{2}},  \quad
q = \frac{Qe}{\epsilon ak_{B}T},\quad \mathrm{Da}=\frac{J_{s}a}{DC_{\infty }}.
\end{equation}
Here, the Damk\"ohler number $\mathrm{Da}$ represents the ratio of the surface reaction rate $J_{s}$ to the diffusive transport rate \citep{moran2017}.

At a given bulk concentration $C_{\infty }$, we prescribe a constant dimensionless surface potential $\psi_s=\psi(1)$ and restrict our consideration to the limit of a thin EDL, $\lambda _D/a \ll 1$. Under these conditions, the fields are decomposed into the leading-order contribution induced by the homogeneous ionic flux and the first-order small perturbations: $c =c^{(0)} +\mathcal{E}c^{(1)}$ and $\psi =\psi ^{(0)} +\mathcal{E}\psi ^{(1)}$. Following the established matched asymptotic expansion framework for the outer ($r\sim 1$) and inner (EDL) regions \citep{asmolov.es:2024,asmolov2024electro}, the leading-order outer solutions take the spherically symmetric forms:
\begin{equation}
c_{o}^{(0)}=1+\frac{\mathrm{Da}}{r},\quad \psi _{o}^{(0)}=-\beta \ln \left( 1+\frac{\mathrm{Da}}{r}\right).
\label{c_sol}
\end{equation}

By using these fields, the generalized particle velocity along the $x$-axis defined by \eqref{rec} can be recast into the dimensionless form:
\begin{equation}
v_{p}=\frac{\mathcal{E}}{6\pi }\int_{V}\left( \widehat{\mathbf{u}}-\mathbf{e}_{x}\right)\cdot \mathbf{f}dV+\frac{2}{3}q\mathcal{E}\times \left\{
\begin{array}{ll}
1, & \text{(electrophoresis)} \\
\beta, & \text{(diffusiophoresis)}
\end{array}\right.,
\label{rec2}
\end{equation}
where $\widehat{\mathbf{u}}\left( \mathbf{r}\right) $ denotes the reference velocity field past a sphere translating at a unit velocity $\mathbf{e}_{x}$ in a stagnant fluid, and $\mathbf{f}=\nabla ^2 \psi ^{(0)}\mathbf{\nabla }\psi ^{(1)}+\nabla ^2 \psi ^{(1)}\mathbf{\nabla }\psi ^{(0)}$ represents the electrical body force density. Finally, evaluating the total dimensionless excess charge of the system from \eqref{qdim} yields the explicit result:
\begin{equation}
q = -\beta \mathrm{Da}.
\label{q_final}
\end{equation}

\begin{figure}[th]
\begin{minipage}{0.48\textwidth}
\centering
\includegraphics[width=1\columnwidth]{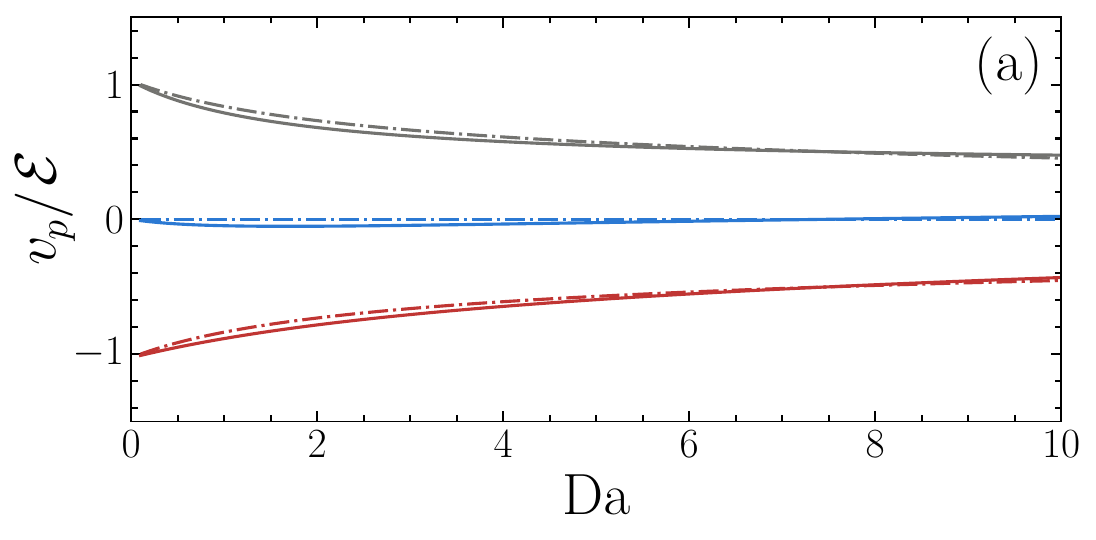}
\end{minipage}
\hfill
\begin{minipage}{0.48\textwidth}
\centering
\includegraphics[width=1\columnwidth]{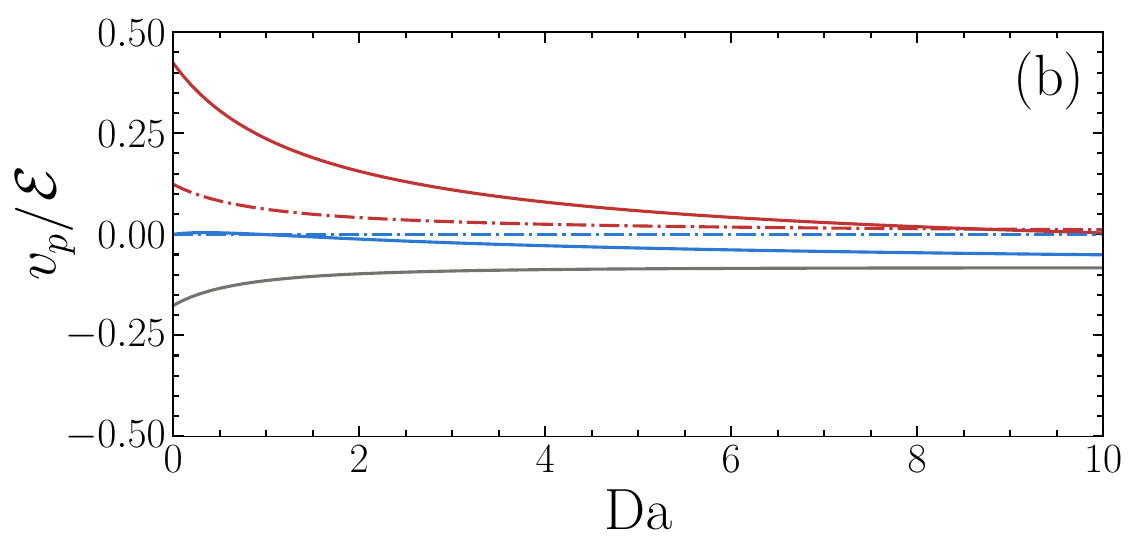}
\end{minipage}
\caption{(a) Electrophoretic mobility of passive catalytic particles vs. $\mathrm{Da}$. The solid and dash-dotted curves show results obtained using $\beta = -0.3$ and $\beta = 0$, respectively. From top to bottom, the surface potentials are $\psi_{s}=1$, $0$, and $-1$. (b) The same for diffusiophoretic mobility, but now the solid curves from top to bottom refer to $\psi_{s}=-1$, $0$, and $1$; the upper dash-dotted curve is evaluated for $\psi_{s}=\pm 1$, and the lower one for $\psi_{s}=0$.}
\label{fig:net}
\end{figure}

Figure~\ref{fig:net} shows the electro- and diffusiophoretic mobilities, $v_{p}/\mathcal{E}$, evaluated analytically from relation \eqref{rec2}. The theoretical predictions obtained using $\beta =-0.3$ (a value typical for inorganic lithium salts) under three different surface potentials are compared directly with reference cases (dash-dotted curves) where the diffusivity contrast is set to zero ($\beta = 0$), implying that no charged secondary cloud emerges. We observe that in all considered cases, the absolute values of both phoretic mobilities systematically decrease with increasing $\mathrm{Da}$.

The results on electrophoretic mobility presented in Fig.~\ref{fig:net}(a) show a quite small discrepancy, which is hardly visible on this scale, between the curves obtained for the two values of $\beta$. This implies that the secondary cloud does not practically affect the electrophoresis of catalytic particles. For inert particles ($\mathrm{Da}=0$), we recover the classical Smoluchowski formula, $v_{p}/\mathcal{E}=\psi_{s}$. If we increase $\mathrm{Da}$, the mobilities of charged particles decay slowly since the disturbance field $-\mathbf{\nabla }\psi^{(1)}$ near the particle reduces, thanks to the enhanced concentration given by Eq.~\eqref{c_sol}. Uncharged particles do not migrate if $\beta = 0$, but when $\beta = -0.3$ their mobility becomes non-zero, although very small, and reverses its sign on increasing $\mathrm{Da}$.

The diffusiophoretic mobility shown in Fig.~\ref{fig:net}(b) is different: there is a significant discrepancy between the mobilities calculated for $\beta =-0.3$ and $\beta =0$. Its nature is apparent. The external electric field is now proportional to $\beta$ (see Eq.~\eqref{rec2}). Thus, it disappears when $\beta =0$, so the migration is of chemiphoretic origin solely. Since the chemiosmotic contribution is an odd function of $\psi _{s}$ \citep{anderson.jl:1989}, the (positive) mobilities calculated for $\psi _{s}=1$ and $\psi _{s}=-1$ coincide. We emphasize that with $\beta =-0.3$ and $\psi _{s}=1$ the mobility becomes negative. This points strongly to the fact that the appearance of the secondary cloud acts not just to change its magnitude, but also its sign.

Our theory can be applied to long-range interactions, where the distance $H$
between the two particles is large compared to their radius ($H \gg a$).
In this case, the concentration and potential fields induced by one particle
near the other act as small disturbances to the self-induced fields, allowing
their gradients to be treated as locally homogeneous. Consequently, the effective
external electric field scales as $E_{\infty} = \frac{Q}{\epsilon H^{2}}$, and
the additional term in Eq.~(\ref{rec}) becomes $E_{\infty}Q = \frac{Q^{2}}{\epsilon H^{2}}$,
with $Q$ given explicitly by (\ref{qdim}). Knowing these interaction fields allows
one to estimate the velocity of the second particle relative to the first due to
the induced phoretic effects.

For instance, the concentration gradient induced by one passive particle near another
can be obtained from~(\ref{c_sol}), which yields $\mathcal{E} = -\mathrm{Da}\left(a/H\right)^{2}$.
Based on this gradient, the relative particle velocities can be readily determined
using the data in Fig.~\ref{fig:net}(b).

\subsection{Microswimmers}

The examples described above correspond to situations of a uniform catalytic particle. In this case, the second term in the reciprocal theorem \eqref{rec} appears only when both $Q$ and $E_{\infty }$ are non-zero. In the case of self-propulsion ($\equiv$ self-diffusiophoresis) of any type of particle \citep{ibrahim2017,de2020self,asmolov2022MDPI,asmolov2022self,xiao2025ionic}, $E_{\infty }=0$ and, hence, $Q$ vanishes as well, which recovers the usual theory. The same holds for swimmers of Type I [Fig.~\ref{fig1}(b)] in an external field \citep{bayati2019}, where the reason for $Q=0$ lies in a zero average flux $\left\langle J\right\rangle$. However, if the swimmers of Type II [Fig.~\ref{fig1}(c)] are subject to an external field, a different story obtains. In this case, both $\left\langle J\right\rangle$ and $E_{\infty }$ are non-zero. Consequently, the extra term in the reciprocal theorem comes into play. Note that the applied field may also tune both the swimmer orientation and the direction of its motion \citep{das2015}. We postpone a detailed analysis of these phenomena to a future paper.

Our results also have repercussions for understanding the electrostatic interaction of two microswimmers, since the field induced by one would appear as external to another. The theory of this phenomenon remains in its infancy, and to our knowledge, only hydrodynamic and chemical interactions have previously been reported \citep{rojas2021,liebchen2021interactions,zottl2023}. It would be of great interest to employ our approach to tackle interactions involving catalytic particles that release ions. The electrostatic interactions for the microswimmers of Type I are negligible due to $Q=0$.
Conversely, for Type II microswimmers, these interactions become significant. In this case,
the swimmers experience long-range interactions analogous to those of passive particles;
however, to properly evaluate their resulting velocities, one must account for the continuous
reorientation of the particles within the induced electric potential and concentration fields.
Investigating these coupled alignment dynamics presents an intriguing direction for future studies.

\section{Conclusion}\label{sec:conclusion}

In this work, we have demonstrated that the traditional paradigm of local
electroneutrality fails for spherical catalytic particles releasing ions with
distinct diffusivities. The initial mismatch in outward diffusion fluxes leads
to the rapid escape of faster ion species into the far-field bulk, leaving
behind an uncompensated net macroscopic excess charge~$Q$. This charge
separation drives the spontaneous formation of a weakly charged secondary
diffuse cloud extending to distances on the order of the particle radius.
By generalizing the Lorentz reciprocal theorem approach, we have established
a phoretic velocity relation that explicitly incorporates the
previously overlooked Coulomb force exerted by the electric field on this
secondary cloud.

These results can be extended from simple spheres to particles of arbitrary
non-spherical shapes. In such generalized configurations, particle translation
and rotation become coupled, meaning that the standard scalar drag is replaced
by the full anisotropic resistance matrix \citep{happel2012low}. Because
the reciprocal theorem naturally accommodates complex boundaries through surface
integration, our framework remains inherently robust; one simply needs to
project the localized electro-hydrodynamic body forces onto the corresponding
components of the reference resistance tensor. This universality ensures that
the discovered phoretic contribution can be evaluated for arbitrary particle
geometries.

Furthermore, a valuable direction for extending the present framework concerns
the regime of finite and large P\'eclet numbers ($Pe \gtrsim 1$). Our current
analytical description is developed under the standard assumption of small
chemical P\'eclet numbers ($Pe \ll 1$), where convective ion transport is
neglected. At higher convective rates, however, the nonlinear coupling
between the surface osmotic flows and solute advection becomes dominant.
As demonstrated in the seminal work by \citet{michelin2013}, this nonlinearity
can trigger a spontaneous fore-aft symmetry breaking of the concentration field,
initiating autophoretic motion even for chemically homogeneous particles.
Incorporating our macroscopic excess charge framework into such convective
advection-dominated regimes represents an important, challenging horizon
for future electrokinetic models. Another fruitful direction could be to extend the analysis of electrophoretic
migration under external electric fields of arbitrary strength~\citep{schnitzer2012}
to ion-releasing particles, where evaluating the total system charge
requires accounting for nonlinear effects.

Finally, our findings open up a fascinating perspective for understanding the long-range electrostatic coupling between multiple active microswimmers. While traditional models of active matter focus exclusively on long-range hydrodynamic and chemical interactions, the self-induced phoretic force discovered here decays slowly as $1/H^2$. For Type II swimmers, this introduces a robust, long-range electrostatic interaction that remains active even at large separation distances. This framework therefore provides a crucial, long-sought physical bridge from single-particle phoretic dynamics to the emergent collective motion, swarming, and phase transitions in active matter systems.

\begin{acknowledgments}

This work was supported by the Ministry of Science and Higher Education of the Russian Federation. The authors declare no competing interests.
\end{acknowledgments}

Declaration of Interests. The authors report no conflict of interest.

\end{document}